\documentclass{sig-alternate-05-2015}
\usepackage{algorithm}
\usepackage{algpseudocode}
\usepackage{hyperref}
\usepackage{tikz}
\usepackage{pgfplots}

\setcopyright{rightsretained}

\toappear{The present work is currently under review by an international conference.}
\title{Improving PageRank for Local Community Detection}

\numberofauthors{3}
\author{
\alignauthor Alexandre Hollocou\\
 \affaddr{INRIA}\\
 \affaddr{Paris, France}\\
 \email{alexandre.hollocou@inria.fr}
\alignauthor Thomas Bonald\\
 \affaddr{Telecom Paristech}\\
 \affaddr{Paris, France}\\
 \email{thomas.bonald\\@telecom-paristech.fr}
\alignauthor Marc Lelarge\\
 \affaddr{INRIA-ENS}\\
 \affaddr{Paris, France}\\
 \email{marc.lelarge@ens.fr}
}

\begin{document}
\maketitle
\begin{abstract}
Community detection is a classical problem in the field of graph mining.
While most algorithms work on the entire graph, it is often interesting
in practice to recover only the community containing some given set of
\emph{seed} nodes. In this paper, we propose a novel approach to this problem, using some
low-dimensional embedding of the graph based on random walks starting
from the seed nodes. From this embedding, we propose some simple
yet efficient versions   of the {PageRank} algorithm as well as
a novel algorithm, called {WalkSCAN}, that is able to
detect  multiple communities, possibly overlapping. We provide insights into the
performance of these algorithms through the theoretical analysis of a
toy network and show that {WalkSCAN} outperforms
existing algorithms on real networks.
\end{abstract}


\section{Introduction}\label{introduction}

\begin{figure*}[ht]
\centering
\includegraphics[height=2in]{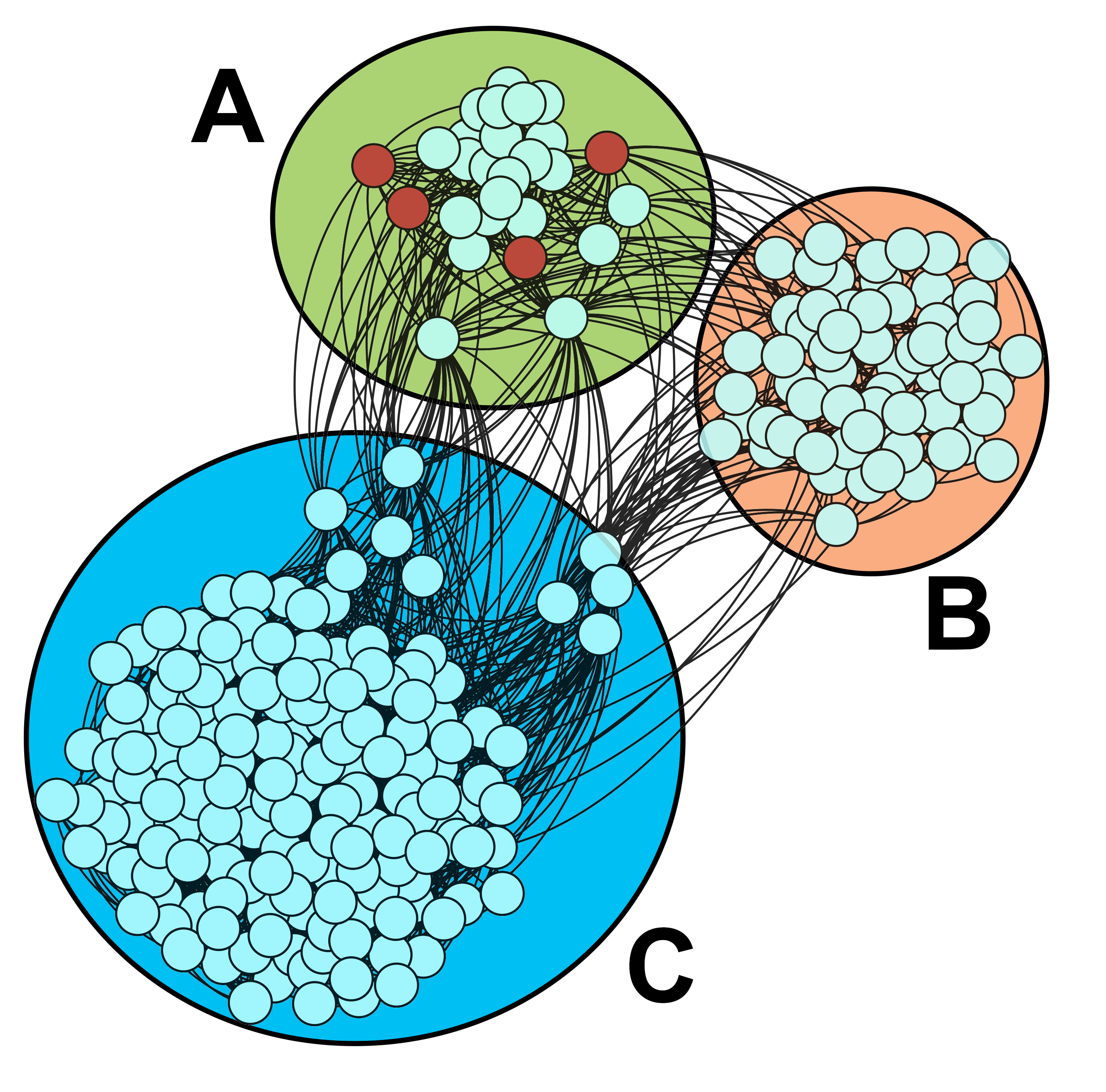}
\hspace{1in}
\includegraphics[height=2in]{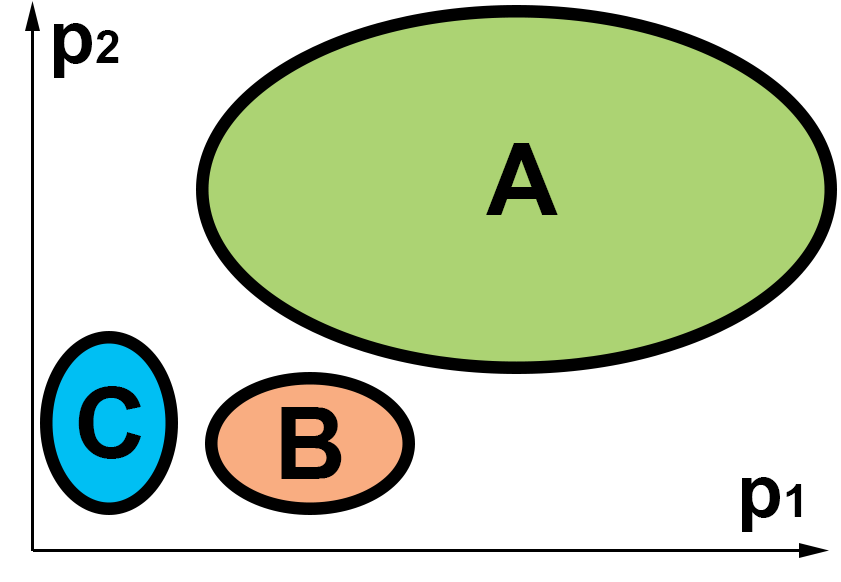}
\caption{Illustration of the random walk embedding with $T=2$.
The left figure shows a network with three communities $A$, $B$ and $C$,
where four seed nodes were picked in $A$ (nodes in red).
The right figure illustrates the result of the embedding
using random walks of length $2$ starting from the seed nodes.
We see that nodes are mapped to three distinct zones in the embedding space
that correspond to communities $A$, $B$ and $C$.}
\label{embedding-illustration}
\end{figure*}

Community detection is a fundamental problem in the field of graph
mining {[}14{]}, with applications to the analysis of social,
information or biological networks. The objective is to find dense
clusters of nodes, the underlying \emph{communities} of the graph. While
most existing algorithms work on the entire graph
{[}15{]}{[}3{]}{[}17{]}{[}11{]}, it is often irrelevant in practice to
cluster all nodes. A more practically interesting problem is to detect
the community to which a given set of nodes, the so-called \emph{seed set}, belong.
This problem, known as \emph{local community detection}
or \emph{seed set expansion}, is particularly relevant in large graphs
where the exploration of the whole network is computationally expensive,
if not impossible.
It is also motivated by the recent empirical results
in {[}7{]}, showing that the community structure of real networks is much
more intricate than a simple partition, as implicitly assumed by most  algorithms working on the entire graph.

The problem of local community detection consists in recovering some unknown community $C$ in a graph given some subset \(S\subset C\) of
these nodes, the seed set, which may be provided by an expert for instance {[}9{]}. A simple approach to this problem is
to add at each step the most important node according to some goodness
metric {[}5{]}{[}10{]}{[}4{]}{[}13{]}. Another class of algorithms use
random walks starting from the seed set
{[}9{]}{[}1{]}{[}8{]}{[}12{]}. For instance, the classical
PageRank algorithm can be combined with some stopping
criterion based on a goodness metric, like conductance or modularity
{[}19{]}{[}18{]}{[}2{]}.

In the present paper, we propose a novel approach to this problem. Specifically, we use
a random walk of $T$ steps starting from the seed set \(S\)  to
embed  the graph in a vector space of  dimension $T$, for some $T\ge 2$.
We notice that the PageRank algorithm can  simply be viewed as a linear classifier in this vector space.
In practice, no more than 3 steps are sufficient to  attain most of
PageRank's power on real networks {[}9{]}, which suggests that the vector space can be of low dimension, say $T=2$ or $3$.
Based on these observations, we
propose some simple yet efficient  versions of PageRank, like
a parameter-free version called LexRank,  consisting
in ranking nodes according to the lexicographical order in the embedding
space rather than to  their PageRank values. Our main contribution is  a novel algorithm, called WalkSCAN,
that clusters the nodes in the embedding space using the popular DBSCAN
algorithm {[}6{]}.

A key feature of WalkSCAN is to output multiple communities, possibly overlapping.
If the seed set $S$ spreads over  multiple communities,
say \(\cup_{i\in I} C_i\), the clustering algorithm allows WalkSCAN
to recover these different communities \((C_i)_{i\in I}\), provided there are enough seed nodes in each of these communities.
This is the main reason why  WalkSCAN outperforms
existing algorithms like PageRank on real networks, whose community structure is very complex {[}16{]}{[}7{]}.
Moreover, our numerical results show that WalkSCAN is robust against errors, in the sense
that it is generally able to recover the target community \(C\) even if the seed
set \(S\) contains some nodes outside \(C\).

The rest of the paper is organized as follows. In the next section, we
describe the graph embedding based on random walks and the resulting
algorithms, including LexRank and WalkSCAN. Some insights into the
performance of these algorithms are provided in Section
\ref{section-algorithm-analysis} through the theoretical analysis of the
embedding on a toy network. The numerical experiments on real
networks are presented in Section \ref{experimental-results}. Section
\ref{section-conclusion} concludes the paper.

\vspace{1.5cm}

\section{Algorithms}\label{algorithms}

\label{section-local-community-detection}

In this section, we first describe our graph embedding  based on random
walks starting from the seed set.
We then use this embedding to revisit the properties of the {PageRank}
algorithm and  to  propose two simple variants of this
algorithm, namely {PageRankThreshold}  and {LexRank}, as well as a novel algorithm, called {WalkSCAN}.

\subsection{Graph embedding}\label{random-walk-local-embedding}

Let \(G=(V,E)\) be an undirected graph. Nodes are assumed to be organized in communities,
in the sense that nodes belonging to the same community tend to be more connected than with the rest of the graph {[}14{]}.
The communities are  unknown and the objective is
to recover a target community \(C\)  given some  seed set $S\subset C$. We denote by $\hat C$ the output set.

A classical approach to this problem consists in  using  random walks starting from the seed set $S$.
Specifically, we choose a  node in $S$ uniformly at random and walk in the graph choosing a neighbor of the current node uniformly at random.
We use \(p_t(v)\) to denote the probability for the walk to be in node
\(v\) at time \(t\). The vector \(p\) is initialized with
\(p_0(v) = 1_S(v) / |S|\), where \(1_S(v)=1\) if \(v \in S\) and
0  otherwise. Then, we have at any time  \(t\):
\[p_{t+1}(v) = \sum_{u\in V} \frac{A_{u,v}}{d(u)}p_t(u),\] where \(A\)
denotes the adjacency matrix of the graph and  \(d(u)=\sum_v A_{u,v}\) the
degree of node \(u\).

From these probabilities, we define a simple embedding that maps each
node of \(v\in V\) to its random walk probabilities up to some specified
time horizon \(T\):
\[v \in V \mapsto \mathbf{p}(v) = (p_1(v),...,p_T(v)) \in [0,1]^T\]
The time horizon can be made very short to limit the exploration of the graph, say $T=2$ or $3$.
It turns out that the graph embedding is meaningful  even in this case.
Taking $T=2$ for instance,  the respective
probabilities for some node $v$ to be reached after one and two steps  should be much higher if $v\in C$. Thus
we expect the
different communities around the seed nodes to be mapped to distinct
zones of the embedding space \([0,1]^2\), as illustrated by Figure \ref{embedding-illustration}.

By clustering the points in the embedding
space, we should be able to recover the target community $C$.
This is the idea of our algorithm {WalkSCAN}, described below.
Before that,
we revisit {PageRank} by interpreting it as a simple linear classifier
in the embedding space and propose some simple yet efficient versions of this  algorithm.

\subsection{Understanding PageRank}\label{understanding-and-improving-pagerank}

\subsubsection*{The PageRank algorithm}\label{pagerank}
\addcontentsline{toc}{subsubsection}{PageRank}

The {PageRank} algorithm is a classical algorithm
for local community detection {[}9{]}{[}21{]}.
It is based on a random walk similar to that described above except that it goes to a neighbor of the current node with probability $\alpha$ and to a node chosen uniformly at random in the seed set $S$
with probability $1-\alpha$. The parameter  $\alpha$ is known as  the {\it damping factor} and is typically set to $0.85$.
This algorithm is sometimes referred to as {\it personalized} PageRank to make the difference with the original {PageRank}  algorithm, corresponding to the case $S=V$.
We use the simple name {PageRank}, as it is common in the literature.

Let   \(r_t(v)\) be  the probability for the walk to be in node
\(v\) at time \(t\). The vector \(r\) is initialized by
\(r_0(v) = 1_S(v) / |S|\) as before but now
$$
r_{t+1}(v) = (1-\alpha) r_0(v) + \alpha \sum_{u\in V} \frac{A_{u,v}}{d(u)}r_t(u).
$$
The  PageRank algorithm ranks the nodes \(v\) by decreasing value of
\(r_T(v)\), for some given time horizon $T$. Let  $v_k$ the  node of rank $k$. A collection of
sets is defined by \(S_k =  \{v_1,\ldots,v_k\}\), for all $k$ such that $r_T(v_k)>0$.
The algorithm then returns the best set \(S\cup S_{k}\) according to some
scoring function \(f\), that is the set $\hat C=S\cup S_{k^\star}$ with
\[k^\star = \arg\max_{k} f(S\cup S_k).\]
This last step is
called the {\it sweep} algorithm. Many different objective functions \(f\) can
be used for this purpose {[}21{]}, the most popular function being
conductance {[}1{]}.
Note  that some algorithms order nodes by
decreasing value of \(r_T(v)/d(v)\) instead of \(r_T(v)$. Theoretical guarantees are then
available for the sweep  algorithm {[}2{]} but the practical relevance of this modified version of PageRank
has recently been questioned {[}9{]}.

\subsubsection*{PageRank in the embedding space}\label{link-with-the-random-walk-embedding}
\addcontentsline{toc}{subsubsection}{Link with the random walk
embedding}

It is straightforward to verify that the PageRank  \(r_T(v)\) of any node $v\not \in S$ can be written
$$ 
r_T(v)=\sum_{t=0}^{T-1} (1 - \alpha)  \alpha^{ t} p_t(v)+\alpha^T p_T(v).
$$
Thus the
{PageRank} algorithm has a direct interpretation in the embedding
space; it returns the nodes \(v\) that are above the hyperplane of
equation:
$$
\sum_{t=1}^{T-1} (1 - \alpha)\alpha^{t} p_t(v) +\alpha^T p_T(v)= \lambda,
$$
for a parameter \(\lambda\) that maximizes the score function f.

Considering the  example of Figure
\ref{embedding-illustration} with \(T=2\), we see that, for some suitable values for  the parameters \(\alpha\) and \(\lambda\), we are able to
recover the target community  by taking nodes above the
line of equation:
$$(1-\alpha)p_1 + \alpha p_2= \frac{\lambda}{\alpha}.$$
Note that the damping factor $\alpha$ characterizes the slope of this line. In particular, it allows one to control the respective weights of $p_1$ and $p_2$ in the PageRank value of each node: when \(\alpha\to 0\), the line is vertical and only
\(p_1\) is used whereas when \(\alpha\to 1\), the
line is horizontal only \(p_2\) is used.
The default value \(\alpha=0.85\) gives $\alpha/(1-\alpha)\approx 5.7$  more
weight to \(p_2\) than to \(p_1\). The parameter \(\lambda\) depends on the local structure of the graph around the seed set $S$  since it is chosen so as to maximize some score function, like
the conductance.

\subsection{Simplifying PageRank}

Given the above observations, it is natural to propose the following simple variants of the PageRank algorithm.

\subsubsection*{PageRankThreshold}\label{pagerankthreshold}
\addcontentsline{toc}{subsubsection}{PageRankThreshold}

The first idea is to use some fixed threshold $\lambda$, which becomes a parameter of the algorithm. We refer to this algorithm as PageRankThreshold.

\begin{algorithm}
\caption{{PageRankThreshold}}
\label{pagerankthreshold}
\begin{algorithmic}[1]
\Require Graph $G=(V,E)$, seed set $S\subset V$, $T,\alpha,\lambda$.
\State Compute the PageRank value $r_T(v)$ of each node $v$.
\State \Return $\hat{C} = S \cup \{v: r_T(v) > \lambda\}$.
\end{algorithmic}
\end{algorithm}

PageRankThreshold is much faster than PageRank, as it does not require to compute the score $f(S\cup S_k)$ for each $k$ such that $r_T (v_k) > 0$, the complexity of the score function being typically of the order of the total degree of the considered set.
Of course, this leaves open the question of the choice of  \(\lambda\). However,
in a semi-supervised learning setting where we have access to some
ground-truth communities in the network, the optimal value for
\(\lambda\) can be computed based on these ground-truth communities and
used in the rest of the network. We  implement this idea in Section
\ref{experimental-results}.

\subsubsection*{LexRank}\label{lexrank}
\addcontentsline{toc}{subsubsection}{LexRank}

The second idea consists in  keeping the objective
function \(f\) but removing the {damping factor} \(\alpha\) in
order to make the {PageRank} algorithm parameter free (except for the time horizon $T$, typically set to $T=2$ or 3). We then need
to find a ranking of the nodes  in order to apply the sweep
algorithm. Recall that in {PageRank} with $T=2$ and $\alpha=0.85$, the sweep algorithm boils down to a linear separation of the points \((p_1(v),p_2(v))\), where
\(p_2(v)\) is given more weight than \(p_1(v)\). It can be argued that $p_1(v)$ (positive for neighbors of the seed nodes) should actually be given more weight than $p_2(v)$. We give it the highest weight by applying the lexicographical
order\footnote{$u\prec v$ for this order if  for some $s \in \{1, ..., T\}$, $p_t(u)=p_t(v)$ for all $t < s$ and $p_s(u) < p_s(v)$.}
in the embedding space.
We refer to this algorithm as LexRank.

\begin{algorithm}
\caption{{LexRank}}
\label{lexrank}
\begin{algorithmic}[1]
\Require Graph $G=(V,E)$, seed set $S\subset V$, $T$, objective function $f$.
\State Compute the random walk vector $\mathbf{p}(v)$ of each node $v$.
\State Rank nodes $v$ by decreasing value of $\mathbf{p}(v)$
in the lexicographical order.
Denote by $v_k$ the node of rank $k$.
\State \Return $\hat{C} = S\cup S_{k^\star}$ with $S_k= \{v_1,\ldots,v_k\}$ and $k^\star =\arg \max_k f(S\cup S_k)$.
\end{algorithmic}
\end{algorithm}

In words,  {LexRank} first compares  the probabilities \(p_1(v)\) to be
at one step from the seed set, then the probabilities \(p_2(v)\) to be
at two steps from the seed set, and so on, whereas {PageRank} compares  linear combinations of these probabilities, with weights depending on the damping factor.

\subsection{WalkSCAN}\label{sec:walkscan}

We now define our local
community detection algorithm, called {WalkSCAN}, which fully exploits
the graph embedding \(v \mapsto \mathbf{p}(v)\) defined
above. A clear limitation of {PageRank}  is the use of a linear classifier to  cluster nodes in the embedding space.
The idea of {WalkSCAN} is to
group  nodes that are close to each other in the embedding
space. In particular, WalkSCAN may output multiple communities, possibly overlapping, and thus reveal the complex  community structure of the graph. These communities are considered in reverse lexicographical order on their mean value in the embedding space, the first community in this order being the more relevant.

To be more precise, let $K_1,\ldots,K_J$ be the connected components of size at least \(2\) in the graph \(G'=(V',E')\) defined by:
\[\begin{split}
V' &=\{ v\in V: \mathbf{p}(v)\neq 0 \},\\
E' &=\{ (u,v) \in V'\times V' : \lVert \mathbf{p}(u) - \mathbf{p}(v) \rVert \leq d \},
\end{split}\] where \(d> 0\) is a distance parameter, and
\(\lVert \cdot \rVert\) is the euclidean norm. Hence, two nodes \(u\)
and \(v\) are in the same connected component \(K_j\) if and only if  there exists a path
between \(u\) and \(v\) of points that are within distance \(d\) of each
other in the embedding space. We also define the set of \emph{outliers}
\(\mathcal{O}\) as the set of nodes of \(V'\) that do not belong to any
set \(K_1,\ldots,K_J\). Note that this approach is a special case of the
popular clustering algorithm {DBSCAN} {[}7{]} with parameters
\(\epsilon = d\) and minSize=2.

Each component \(K_j\) is the  \emph{core} of some community
\(\hat{C}_j\), which  is obtained from  \(K_j\)
by adding the outliers having at least one neighbor in \(K_j\) in the
original graph \(G\). Note  that whereas the sets \(K_j\) are
disjoints, an outlier can have   neighbors in   different community
cores, so that WalkSCAN may output   overlapping communities.

\begin{algorithm}
\caption{WalkSCAN}
\label{walkscan}
\begin{algorithmic}[1]
\Require Graph $G=(V,E)$, seed set $S\subset V$,  $T,d$.
\State Compute the random walk vector $\mathbf{p}(v)$ of each node $v$.
\State Create the graph $G'=(V', E')$ with:\newline
$V' =\{ v\in V: \mathbf{p}(v)\neq 0 \}$\newline
$E' =\{ (u,v) \in V'\times V' : \lVert \mathbf{p}(u) - \mathbf{p}(v) \rVert \leq d \}$
\State Compute the connected components of $G'$.
Let $K_1,\ldots,K_J$ be the components of size $\geq 2$ and $\mathcal{O}$ the set of isolated nodes in $V'$.
\State Add neighboring nodes from $\mathcal{O}$ to each community:
$\hat{C}_j \leftarrow K_j \cup (\text{Nei}_G(K_j) \cap \mathcal{O})$ for all $j$.
\State Compute $\hat{\mathbf{p}}_j=\frac 1 {|\hat{C}_j|}\sum_{v\in \hat{C}_j} \mathbf{p}(v)$ for all $j$.
\State \Return $\hat{C}_1,\ldots,\hat{C}_J$ in reverse  lexicographical order of $\hat{\mathbf{p}}_1,\ldots,\hat{\mathbf{p}}_J$.
\end{algorithmic}
\end{algorithm}

\section{Analysis}\label{analysis-of-the-embedding}

\label{section-algorithm-analysis} In this section, we show on a toy  network
consisting of overlapping cliques that clusters of nodes in the
embedding space correspond to these cliques, even in the simplest case where $T=2$. We demonstrate that this
property still holds when we add noise to the model, and deduce some insights into the behavior of our  algorithms. Then, we
qualitatively compare the results obtained in the analysis to
observations in a  real-world network.

\subsection{Embedding of a toy network}\label{analysis-on-a-simple-model}

\subsubsection*{Overlapping cliques}\label{overlapping-cliques}
\addcontentsline{toc}{subsubsection}{Overlapping cliques}

\label{section-overlapping-cliques}

First, we consider the case of two overlapping cliques isolated from the
rest of the graph. Let \(V\) be a set of nodes, and \(C_1\) and \(C_2\)
be two subsets  of \(V\) such that \(C_1 \cap C_2 \neq \emptyset\). We assume
that \(C_1\) and \(C_2\) form cliques and are
disconnected from \(V\setminus (C_1 \cup C_2)\). The adjacency matrix
\(A\) of the graph satisfies: \[A_{u,v} =
\begin{cases}
1 & \text{if }u, v \in C_1 \text{ or } u, v \in C_2, \\
0 & \text{otherwise}.
\end{cases}\]

Assume that \(C_1\) is the target community and that \(S \subset C_1\), with $S\cap (C_1 \setminus C_2) \ne \emptyset$.
We take $T=2$ for simplicity; the analysis can be easily extended to the case $T\ge 3$.
We show in Appendix \ref{appendix-overlapping-cliques} that there exists
three vectors \(\mathbf{p}_{1\cap 2}\), \(\mathbf{p}_{1\setminus 2}\)
and \(\mathbf{p}_{2\setminus 1}\) in \(\mathbb{R}^2\) such that, for all
\(v\in V \setminus S\), \[ \mathbf{p}(v) =
\begin{cases}
\mathbf{p}_{1 \cap 2} & \text{if }v \in C_1\cap C_2, \\
\mathbf{p}_{1\setminus 2} & \text{if }v \in C_1 \setminus C_2, \\
\mathbf{p}_{2\setminus 1} & \text{if }v \in C_2 \setminus C_1, \\
0 & \text{otherwise}.
\end{cases}
\]
These vectors satisfy $\mathbf{p}_{1 \cap 2} \ge  \mathbf{p}_{1\setminus 2}, \mathbf{p}_{2\setminus 1}$ componentwise so that both PageRank and PageRankThreshold would recover
either $C_1 \cap C_2$, $C_1$ or $C_2$, or $C_1\cup C_2$, depending on the sweep algorithm or the threshold $\lambda$, the output $C_1$ or $C_2$ depending on the damping factor $\alpha$ and the respective sizes of these communities compared to their intersection.
Since $\mathbf{p}_{1 \cap 2} \succ \mathbf{p}_{1\setminus 2} \succ \mathbf{p}_{2\setminus 1}$ for the lexicographical order, LexRank will always  recover either $C_1 \cap C_2$, $C_1$ or $C_1\cup C_2$,
depending on the sweep algorithm, which suggests that this algorithm is less sensitive to the respective sizes of the communities.

Now defining
\begin{align}
d_1 &= \lVert \mathbf{p}_{1\setminus 2} - \mathbf{p}_{1 \cap 2} \rVert, \label{eq:d1}\\
d_2 &= \min(\lVert \mathbf{p}_{2\setminus 1} - \mathbf{p}_{1 \setminus 2} \rVert, \lVert \mathbf{p}_{2 \setminus 1} - \mathbf{p}_{1 \cap 2} \rVert ),\label{eq:d2}
\end{align}
we have
\(d_1 < d_2\) if \(|S \cap C_1 \cap C_2|\) is small enough compared to
\(|S \cap (C_1 \setminus C_2)|\) (see Appendix
\ref{appendix-overlapping-cliques}).
If the distance parameter of WalkSCAN  satisfies
\(d < d_1 < d_2\), we have three core sets  corresponding to \(C_1 \cap C_2\), \(C_1 \setminus C_2\) and
\(C_2 \setminus C_1\), and the two communities $C_1$ and $C_2$ can be fully recovered:
WalkSCAN  outputs $C_1 \cap C_2$, $C_1\setminus C_2$ and $C_2\setminus C_1$ in this order,
which allows to recover both communitites $C_1$, $C_2$ and their intersection.
If \(d_1 < d < d_2\), we
have two core sets   corresponding to \(C_1\) and
\(C_2 \setminus C_1\) and WalkSCAN will output $C_1$ and $C_2\setminus C_1$ in this order. Next, we study a
more realistic model obtained by removing edges in the cliques or by
adding  edges between \(C_1 \cup C_2\) and
\(V \setminus (C_1 \cup C_2)\).

\subsubsection*{From cliques to quasi-cliques}\label{from-cliques-to-quasi-cliques}
\addcontentsline{toc}{subsubsection}{From cliques to quasi-cliques}

Consider a model where   \(k\) edges are removed  from \(C_1\) or
\(C_2\), which is more representative of  real-world networks where communities are
dense sets but not  cliques in general. Assume that $S\subset C_1\setminus C_2$.
Denoting by \(\bar {\mathbf{p}}(v)\) the initial graph embedding, without removing these edges, we show in
Appendix \ref{appendix-noisy-cliques} that
\[
\lVert  \mathbf{p}(v) - \bar {\mathbf{p}}(v) \rVert    \le d_2\] for all nodes $v$ but $k$ in the worst case scenario, provided  \(k\) is small
enough compared to \(|C_1|\).
Thus, if the distance parameter of WalkSCAN satisfies \(d_1 < d < d_2\), where $d_1$ and $d_2$ are defined by  \eqref{eq:d1}-\eqref{eq:d2} (for the cliques),
the algorithm will output \(C_1\) and \(C_2 \setminus C_1\) in this order.

\subsubsection*{Adding external links}\label{adding-external-links}
\addcontentsline{toc}{subsubsection}{Adding external links}

Now consider the two-clique model where $l$ edges are added
between nodes of \(C_1 \cup C_2\) and nodes of
\(V \setminus (C_1 \cup C_2)\).
We show in Appendix \ref{appendix-adding-external-links} that
\begin{align*}
&|| \mathbf{p}(v) - \bar{\mathbf{p}}(v)||< d_2 \ \text{ for }v \in C_1 \cup C_2,\\
& p_2(v)\le \min((\mathbf{p}_{1\cap 2})_2,(\mathbf{p}_{1\setminus 2})_2,(\mathbf{p}_{2\setminus 1})_2)- d \ \text{ otherwise,}
\end{align*}
provided  \(|C_1|/l\) is large enough. Thus, if \(d_1 < d < d_2\), WalkSCAN will output \(C_1\) and \(C_2 \setminus C_1\) in this order.

\subsection{Embedding of a real-world network}\label{real-world-networks}

\begin{figure}[ht]
    \begin{tikzpicture}
    \begin{axis}[
        legend style={at={(0.35,0.95)}, anchor=north,legend columns=2},
        legend cell align=left,
        ticklabel style = {/pgf/number format/fixed},
        xlabel=$p_1$,
        ylabel=$p_2$,
        scatter/classes={
            ab={violet!60!white},
            a={blue!60!white},
            b={red!60!white},
            u={black!60!white}}]
        \addplot[
            scatter,
            only marks,
            scatter src=explicit symbolic,
            visualization depends on=\thisrow{size}\as\msize,
            visualization depends on={\thisrow{size}>1 \as \gtone},
            scatter/@pre marker code/.append style={
                /tikz/mark size=1+\msize
            },
            scatter/@pre marker code/.append style={
                /tikz/mark={\ifdim\gtone pt=1 pt *\else triangle*\fi}
            },
        ]
        table[meta=label] {
    x        y        label   size
    0.0000   0.0069   u       5
    0.0000   0.0200   b       6
    0.0694   0.0594   b       1
    0.0000   0.0525   b       2
    0.1278   0.0864   a       1
    0.0278   0.0900   ab      1
    0.1694   0.1400   ab      1
    0.0778   0.0964   a       2
    0.1694   0.0883   a       1
    0.1417   0.1356   ab      2
    };
    \legend{$C_1 \cap C_2$,$C_1 \setminus C_2$,$C_2 \setminus C_1$,$V \setminus (C_1 \cup C_2)$}
    \end{axis}
\end{tikzpicture}
\caption{Graph embedding for two overlapping communities $C_1, C_2$ of the DBLP dataset.
Disks correspond to sets of  nodes, triangles to isolated nodes. Colors indicate the sets to which each node belongs.}
\label{real-world-embedding-result}
\end{figure}
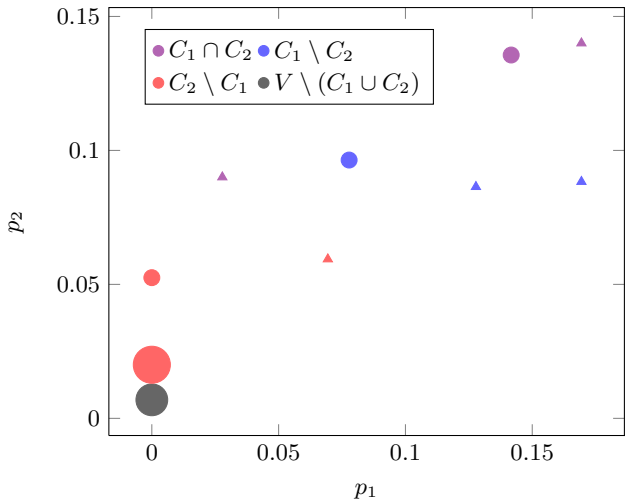

In Figure \ref{real-world-embedding-result}, we plot the embedding \((p_1(v), p_2(v))\) of the DBLP graph
for a seed set \(S\) consisting of two nodes  in a given ground-truth community, say $C_1$  (see Section \ref{experimental-results} for details on the dataset).
One of these nodes belongs to another community $C_2$.
We observe groups of nodes sharing the same feature values
in the embedding space, represented by disks, and  nodes with singular values in the embedding space,
represented by triangles.

The results  correspond qualitatively to the above analysis.
Depending on the sweep algorithm or the threshold $\lambda$, PageRank, PageRankThreshold and LexRank will approximately output either $C_1\cap C_2$, $C_1$ or $C_1\cup C_2$, LexRank being more efficient for identifying $C_2\setminus C_1$.
By clustering  points  in the
embedding space, WalkSCAN is able to recover the complete community structure, namely
 \(C_1\), \(C_2\) and their intersection, up to some isolated points, depending on the distance parameter $d$.
\vspace{1em}

\section{Experimental results}\label{experimental-results}

We now analyse the performance of our algorithms through numerical experiments on real networks.

\label{experimental-results}

\subsection{Benchmark}

\subsubsection*{Datasets}\label{datasets}
\addcontentsline{toc}{subsubsection}{Datasets}

We use two datasets of real networks available from the Stanford Social Network Analysis Project (SNAP)
{[}21{]}: DBLP (graph of co-authorship) and YouTube (online social graph).
These datasets, whose main features are  given in  Table \ref{datasets-characteristics}, include ground-truth community memberships that
we use to measure the quality of the results. We use the top 5000 communities
in each of these datasets for our benchmarks.

\begin{table}[ht]
    \begin{center}
        \begin{tabular}{l ccc}
        \hline
        Dataset     & Nodes     & Edges           & Communities\\
        \hline \hline
        DBLP        & 317,080   & 1,049,866       & 13,477     \\
        \hline
        YouTube     & 1,134,890 & 2,987,624       & 8,385      \\
        \hline
        \end{tabular}
    \end{center}
\caption{Main features of the considered datasets.}
\label{datasets-characteristics}
\end{table}

\subsubsection*{Algorithms}\label{algorithms-1}
\addcontentsline{toc}{subsubsection}{Algorithms}

We perform experiments for {PageRank} (PR), {WalkSCAN} (WS),
{PageRankThreshold} (PRT) and {LexRank} (LR). We use the
conductance as the objective function for both {PageRank} and
{LexRank}.
The threshold parameter $\lambda$ of PageRankThreshold is optimized based on half of the available communities of each dataset, according to the envisaged semi-supervised learning setting for this algorithm.
We have implemented all these algorithms in C++ and made them
available on GitHub\footnote{\url{https://github.com/ahollocou/walkscan}}.

We use the value \(T=3\) for {PageRank}, {LexRank} and
{PageRankThreshold} because we have observed that greater values of \(T\)
do not bring significantly better results. This corroborates the
results of  {[}9{]} concerning PageRank. For {WalkSCAN}, we
take  \(T=2\), which turns out to provide the best results (although the results are very similar for $T=3$).

Note that we do not included  {LEMON} {[}12{]} and {Heat
Kernel} {[}8{]}  in our benchmark because our experiments have shown that they never  outperform {PageRank}.

\subsubsection*{Performance metric}\label{f1-score-evaluation}
\addcontentsline{toc}{subsubsection}{F1 score evaluation}

We use the {F1-Score} to evaluate the performance of the
algorithms. The F1-Score
\(\text{F}1(\hat{C}, C)\) of the output \(\hat{C}\) for the target community \(C\) is the harmonic mean of the
precision and the recall of \(\hat{C}\) with respect to \(C\):
\[\text{F}1(\hat{C}, C) = H(\mathrm{precision}(\hat{C}, C), \mathrm{recall}(\hat{C}, C))\]
where \(H(a,b) = \frac{2 a b}{a + b}\) and
\[\mathrm{precision}(\hat{C}, C) = \frac{|\hat{C} \cap C|}{|\hat{C}|},\quad
\mathrm{recall}(\hat{C}, C) = \frac{|\hat{C} \cap C|}{|C|}.\]

\subsection{Recovering a single
community}\label{recovering-single-communities}

\begin{figure}
    \begin{tikzpicture}
        \begin{axis}[
            symbolic x coords={Amazon,DBLP,YouTube,Orkut},
            xtick=data,
            ylabel=Average F1 Score,
            enlarge x limits=0.6,
            legend style={at={(0.5,0.85)},anchor=west,legend columns=2},
            legend cell align=left,
            ybar,
            bar width=15pt,
            ymin=0,
        ]
            \addplot [fill=cyan]
                coordinates {(DBLP, 0.716) (YouTube, 0.465)};
            \addplot [fill=purple]
                coordinates {(DBLP, 0.740) (YouTube, 0.484)};
            \addplot [fill=orange]
                coordinates {(DBLP, 0.713) (YouTube, 0.467)};
            \addplot [fill=brown!50!white]
                coordinates {(DBLP, 0.726) (YouTube, 0.512)};
            \addplot [fill=brown]
                coordinates {(DBLP, 0.751) (YouTube, 0.526)};
            \legend{PR,PRT,LR,WS,WS-Expert}
        \end{axis}
    \end{tikzpicture}
\caption{Benchmark on real-networks from SNAP.}
\label{benchmark1}
\end{figure}
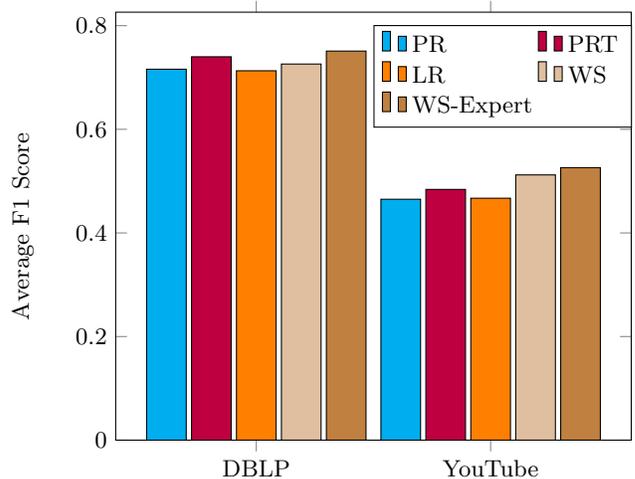

For each dataset and each ground-truth community $C$, we choose a random subset \(S\) of
\(C\) of size \(\lceil |C| / 10 \rceil\) as the seed set and compute the F1 score of each algorithm. The results are averaged over the $N$ ground-truth communities
(recall that we take $N=5000$ for each dataset).

A basic approach to assess the performance of WalkSCAN is to consider only the first community returned by the algorithm. This is a bit restrictive, however, since valuable information can be retrieved from the other communities returned by the algorithm.
If we have access to some experts for instance, one may submit the first $K$ communities  returned by WalkSCAN to these experts
and  let them choose the best one among these.
This is equivalent to computing the maximum of $ \text{F}1(\hat{C}_j, C)$ for the $K$ first communities $\hat C_j$ returned by WalkSCAN. We denote the corresponding algorithm by WS-Expert (WalkSCAN helped by Experts).
In practice, $K$ can be very small. Here, we use $K=2$, and observe that the performance shows pratically no increase for higher values of $K$.

Figure \ref{benchmark1} shows the results for DBLP and YouTube. We notice that both WS and WS-Expert show better results than PR.
In particular, {WS-Expert} outperforms PR by \(13\%\) on YouTube and by \(5\%\) on DBLP.
{PRT} slightly outperforms {PR}
but the results cannot be compared directly since PRT uses some information on the network (to determine the best threshold $\lambda$).
However, the results   suggest that  conductance is not the best objective function for the sweep algorithm, and that a simple threshold on the PageRank values is preferable.
{LR} shows essentially the same performance as {PR},
with  the  advantage of being parameter free.

\subsubsection*{\texorpdfstring{Distance parameter of WalkSCAN}}\label{performance-in-function-of-d}
\addcontentsline{toc}{subsubsection}{Performance in function of \(d\)}

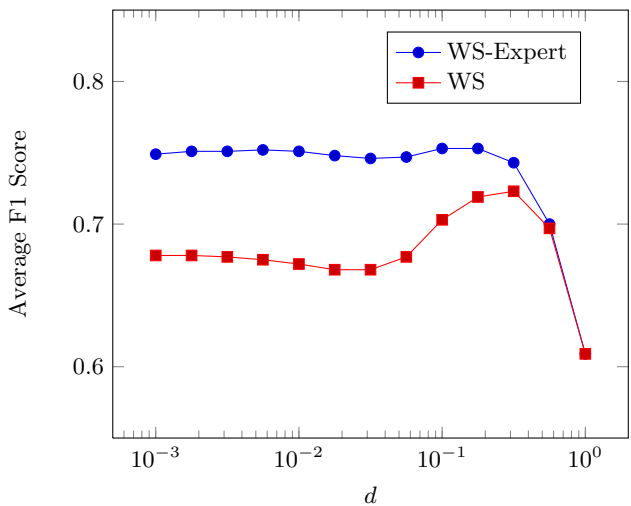
\begin{figure}[ht]
    \begin{tikzpicture}
        \begin{semilogxaxis}[
            xlabel=$d$,
            ylabel=Average F1 Score,
            ymin = 0.55,
            ymax = 0.85,
            legend style={at={(0.72,0.95)},anchor=north,legend columns=1},
            legend cell align=left
        ]
            \addplot
                coordinates {(0.001, 0.749) (0.00178, 0.751) (0.00316, 0.751) (0.0056, 0.752)
                             (0.01, 0.751) (0.01778, 0.748) (0.03162, 0.746) (0.05623, 0.747)
                             (0.1, 0.753) (0.1778, 0.753) (0.316, 0.743) (0.562, 0.700)
                             (1, 0.609)};
            \addplot
                coordinates {(0.001, 0.678) (0.00178, 0.678) (0.00316, 0.677) (0.0056, 0.675)
                             (0.01, 0.672) (0.01778, 0.668) (0.03162, 0.668) (0.05623, 0.677)
                             (0.1, 0.703) (0.1778, 0.719) (0.316, 0.723) (0.562, 0.697)
                             (1, 0.609)};
            \legend{WS-Expert,WS}
        \end{semilogxaxis}
    \end{tikzpicture}
\caption{Impact of the distance parameter $d$ on WalkSCAN performance (DBLP dataset).}
\label{fig:performance-in-function-of-d}
\end{figure}

Figure \ref{fig:performance-in-function-of-d} shows the impact of the distance parameter
\(d\) on the performance of {WalkSCAN}.
We observe that the average detection score of WS-Expert is nearly constant for values of \(d\) lower than \(0.1\),
whereas the performance of WS increases notably around $d = 0.4$.
This behavior is consistent with the results of the analysis from the previous section.
Recall that WS only considers the community $\hat{C}_1$, whereas WS-Expert choses the best
community among $\hat{C}_1,...,\hat{C}_K$.
When $d < d_1 < d_2$ in our simple model, $\hat{C}_1$ corresponds to $C_1 \cap C_2$ and $\hat{C}_2$ to $C_1 \setminus C_2$.
Then WS returns $C_1 \cap C_2$, whereas WS-Expert returns $C_1 \setminus C_2$,
which leads to a better F1 score.
If $d_1 < d < d_2$, $\hat{C}_1$ corresponds to $C_1$,
and both WS and WS-Expert return $C_1$.
This explains why the detection score of WS is more sensitive to the value of $d$ than WS-Expert.

\subsubsection*{Merging communities: WS-Merge}\label{multi-scale-communities-ws2}
\addcontentsline{toc}{subsubsection}{Multi-scale communities: WS2}

\begin{figure}[ht]
    \begin{tikzpicture}
        \begin{axis}[
            symbolic x coords={Amazon,DBLP,YouTube,Orkut},
            xtick=data,
            ylabel=Average F1 Score,
            enlarge x limits=0.6,
            legend style={at={(0.6,0.85)},anchor=west,legend columns=1},
            legend cell align=left,
            ybar,
            bar width=20pt,
            ymin=0,
        ]
            \addplot 
                coordinates {(DBLP, 0.716) (YouTube, 0.465)};
            \addplot 
                coordinates {(DBLP, 0.751) (YouTube, 0.526)};
            \addplot[fill=purple!50!white]
                coordinates {(DBLP, 0.797) (YouTube, 0.530)};
            \legend{PR,WS-Expert,WS-Merge}
        \end{axis}
    \end{tikzpicture}
\caption{Impact of merging communities on the performance of WalkSCAN.}
\label{benchmark2}
\end{figure}
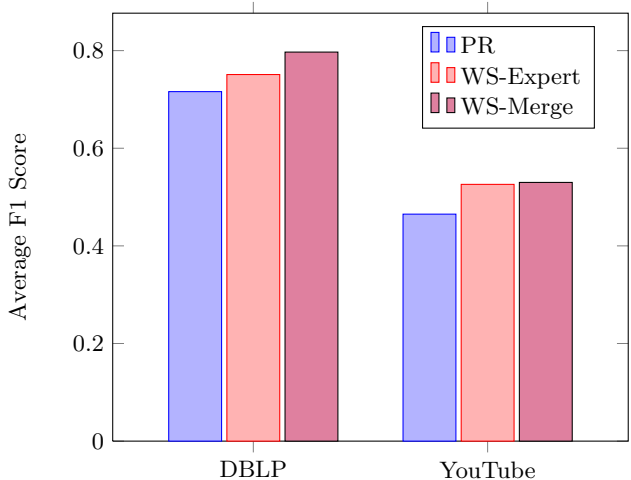

We now study the performance gain achieved when the expert of WS-Expert is allowed to merge two communities.
This is equivalent to  computing the maximum of $ \text{F}1(\hat{C}_j, C)$ and $ \text{F}1(\hat{C}_i\cap \hat C_j, C)$ among the $K$ first communities  returned by WalkSCAN.
We refer to this modified version of Walk- SCAN as WS-Merge.
In the two-clique model of \S\ref{analysis-on-a-simple-model}, WS-Merge would output  $C_1$ whereas WS would output either $C_1\cap C_2$ or $C_1$.
The results given in Figure \ref{benchmark2} show that WS-Merge outperforms WS-Expert (this increase is especially noticeable on DBLP).
In the end, the average F1 score of WS-Merge is \(11\%\)
higher on DBLP, and \(14\%\) higher on YouTube, than the one of
{PageRank}.

This can be explained by the fact that ground-truth communities in the
SNAP datasets can often be decomposed into sub-communities {[}7{]}.
{WalkSCAN} sometimes recovers these sub-communities rather than the
ground-truth community as labeled in the dataset. This raises the important question of the
multi-scale community structure of real networks. There is
no universal definition of a community, just because this
depends implicitely on the  scale adopted. We believe that WalkSCAN is an important step towards the multi-scale analysis of communities in networks.
The following experiments illustrate the ablity of WalkSCAN to detect multiple communities.

\subsection{Recovering  multiple communities}\label{extension-to-multiple-communities}

\label{section-detecting-multiple-communities}

\subsubsection*{A motivating example}\label{a-motivating-example}
\addcontentsline{toc}{subsubsection}{A motivating example}

If our seed set contains nodes from different communities, or if one or
more seed nodes belong simultaneously to several overlapping
communities, standard algorithms like PageRank are generally not
able to recover all these  communities. This is an important
limitation  as in real networks, each community has
many other communities in its neighborhood, and these communities are
very likely to overlap.

If we consider that seed nodes are recommended by domain experts
{[}9{]}, these experts might very well make some mistakes and propose
nodes outside the target community, or recommend nodes that belong to
the target community and a neighboring community at the same time. We
would like to be able to recover the correct community even if we have
such \emph{mixed} seed sets.
In the two-clique model of \S\ref{analysis-on-a-simple-model}, this would correspond to some seed set $S\subset C_1\cup C_2$.
It can be easily checked that the complete community structure ($C_1$, $C_2$ and their intersection) can still be recovered by WalkSCAN
provided the proportion of seed nodes in the intersection is sufficiently small (otherwise, only $C_1\cap C_2$ or $C_1\cup C_2$ is recovered).

In the following benchmarks, we test on real networks the ability of
{WalkSCAN} to recover the correct communities when
the seed nodes belong to more than one community.

\subsubsection*{Random seed set}\label{random-seed-set}
\addcontentsline{toc}{subsubsection}{Random seed set}

\begin{figure}
    \begin{tikzpicture}
        \begin{axis}[
            xlabel=Seed set size $k$,
            ylabel=Average F1 Score,
            ticklabel style = {/pgf/number format/fixed},
            ymax=0.10,
            legend style={at={(0.72,0.95)},anchor=north,legend columns=1},
            legend cell align=left
        ]
            \addplot[color=magenta,mark=*]
                coordinates {(1, 0.0877) (2, 0.0520) (3, 0.0422) (4, 0.0362) (5, 0.0340)};
            \addplot[color=teal,mark=square]
                coordinates {(1, 0.0972) (2, 0.0471) (3, 0.0326) (4, 0.0208) (5, 0.0190)};
            \legend{WS,PR}
        \end{axis}
    \end{tikzpicture}
\caption{Performance of PageRank and WalkSCAN for  random seed sets of size $k$ (DBLP dataset).}
\label{random-seed-set-benchmark}
\end{figure}
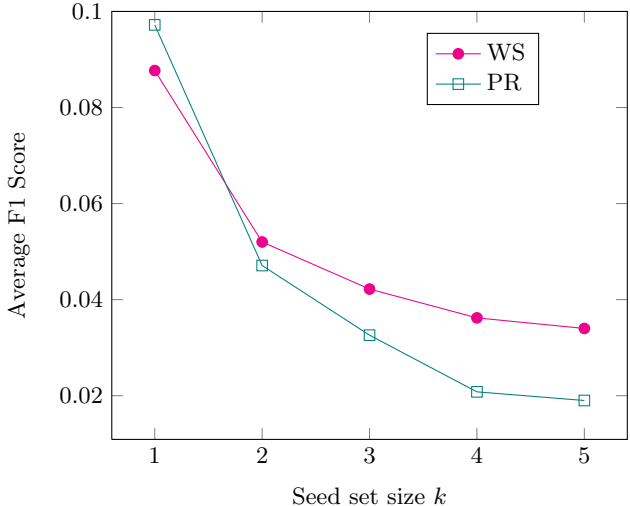

For any given integer \(k\ge 1\), we
take random sets $S\subset V$ of size \(k\) as seed sets.
The target  community is  the union of all communities
to which these seed nodes belong.
We compare the performance of PageRank (with the conductance as objective function) and WalkSCAN,
where the estimated community correponds to the  merging of all the communities $\hat{C}_j$ returned by the algorithm.
The  results are shown in  Figure \ref{random-seed-set-benchmark} for the DBLP dataset and 1000 independent runs per value of $k$. We see that the
performance of PR drops significantly faster than that of WS
as $k$ increases.

\subsubsection*{Locally random seed set}\label{locally-random-seed-set}
\addcontentsline{toc}{subsubsection}{Locally random seed set}

\begin{figure}
    \begin{tikzpicture}[scale=0.95]
        \begin{axis}[
            xlabel=Maximum distance  $l$,
            ylabel=Average F1 Score,
            legend style={at={(0.72,0.95)},anchor=north,legend columns=1},
            legend cell align=left,
        ]
            \addplot[color=magenta,mark=*]
                coordinates {(1, 0.573) (2, 0.340) (3, 0.274)};
            \addplot[color=teal,mark=square]
                coordinates {(1, 0.397) (2, 0.115) (3, 0.029)};
            \legend{WS-Expert,PR}
        \end{axis}
    \end{tikzpicture}
\caption{Performance of PageRank and WalkSCAN for  {locally random} seed sets at maximum distance $l$ to the target community (DBLP dataset).}
\label{locally-random-seed-set-benchmark}
\end{figure}
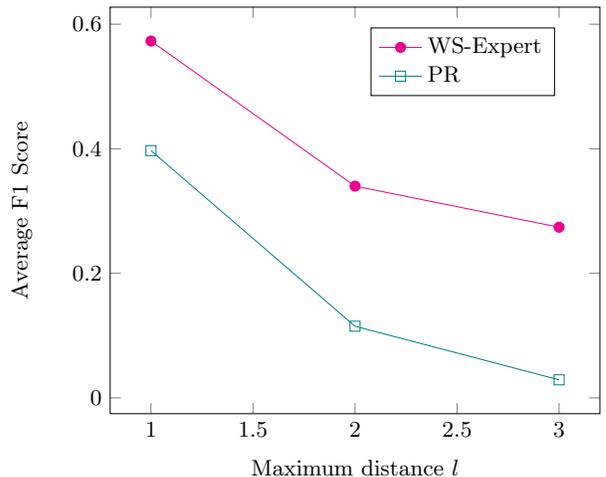

In practice, experts do not choose completely random seed nodes in
\(V\), but nodes around a target community. Here, we consider a more
realistic setup where given some ground-truth community \(C\), we pick
\(\lceil |C|/10\rceil\) nodes at random in \(C\) or at distance at most \(l\) of
\(C\).
The results are averaged over all ground-truth communities of the dataset.
We use WS-Expert to measure the performance of WalkSCAN, which is consistent with the choice of the seed set.
The results are shown in Figure \ref{locally-random-seed-set-benchmark} for  the DBLP dataset. Again, {WS-Expert} outperforms {PR} significantly.

\section{Conclusion}\label{conclusion}

\label{section-conclusion}

In this paper, we have introduced an embedding of the local structure of
the network into a low-dimensional vector space based on random walks
starting from the seed nodes. We have seen both theoretically on a clique-based model  and experimentally on
real-world networks  that this
embedding has very interesting properties for detecting communities
in the neighborhood of the seed set. Indeed, nodes belonging to
different communities in the network are well separated in the embedding
space.

Building on this embedding, we have designed a new algorithm, called
{WalkSCAN}, that outperforms the state-of-the-art {PageRank}
algorithm on real data and is able to recover multiple
communities in the neighborhood of the seed nodes. {WalkSCAN} is
therefore more adapted to real cases where we do not necessarily control
the quality of the seed set, in the sense that seeds are likely to
belong to multiple communities.

The random-walk embedding also gives  insights on the behavior of
the  {PageRank} algorithm, which can be viewed as a linear classifier in the embedding space. From these observations, we have
introduced two variants of {PageRank},
{PageRankThreshold} and {LexRank}, that slightly outperform PageRank on real networks: {PageRankThreshold}
has the advantage not to depend on a computationally expensive
objective function, while {LexRank}  is parameter-free.

This work opens interesting research directions. First, we would like to
study the possibility of using cores  as new seed sets for seed
set expansion algorithms. This re-seeding strategy could be used to
improve {WalkSCAN}, or could be applied to the detection of
overlapping communities. Second, we plan to explore
the potential benefits of adding node attributes or edge attributes as
new dimensions of the embedding to perform community detection on
annotated networks.
Finally, we would like to  explore the interest of our embedding to other contexts like network visualization and link prediction.

\appendix

\section{Overlapping cliques}
\label{appendix-overlapping-cliques}

Consider two overlapping cliques $C_1$ and $C_2$, as described in \S\ref{analysis-on-a-simple-model}.

\subsection*{Single seed node}

We first assume that the seed set is reduced to a single node, say $S=\{v_0\}$.

\subsubsection*{\texorpdfstring{Case
\(S \subset C_1 \setminus C_2\)}{Case S \textbackslash{}subset C\_1 \textbackslash{}setminus C\_2}}\label{case-s-subset-c_1-setminus-c_2}

We have: \[
\begin{split}
p_1(v) &=
\begin{cases}
 \frac{1}{d(v_0)} & \text{if }v \in C_1,\\
0 & \text{otherwise},
\end{cases}\\
p_2(v) &=
\begin{cases}
 \frac{1}{d(v_0)}\sum_{u\in C_1} \frac{1}{d(u)} & \text{if }v \in C_1,\\
 \frac{1}{d(v_0)}\sum_{u\in C_1 \cap C_2} \frac{1}{d(u)} & \text{if }v \in C_2\setminus C_1,\\
0 & \text{otherwise}.
\end{cases}
\end{split}
\]

Thus, if \(\alpha_1 = \frac{|C_1 \setminus C_2|}{|C_1|}\) and
\(\beta = \frac{|C_1 \cap C_2|}{|C_1 \cup C_2|}\): \[ \mathbf{p}(v) =
\begin{cases}
    \frac{1}{|C_1|} (1, \alpha_1 + \beta) \overset{def} {=} \mathbf{p}_1 & \text{if }v \in C_1,\\
\frac{1}{|C_1|} (0, \beta) \overset{def}{=} \mathbf{p}_2 & \text{if }v \in C_2\setminus C_1,\\
0 & \text{otherwise}.
\end{cases}
\]

\subsubsection*{\texorpdfstring{Case
\(S \subset C_1 \cap C_2\)}{Case S \textbackslash{}subset C\_1 \textbackslash{}cap C\_2}}\label{case-s-subset-c_1-cap-c_2}

We have: \[
\begin{split}
p_1(v) &=
\begin{cases}
\frac{1}{d(v_0)} & \text{if }v \in C_1 \cup C_2,\\
0 & \text{otherwise}.
\end{cases}\\
p_2(v) &=
\begin{cases}
\frac{1}{d(v_0)} \sum_{u\in C_1 \cup C_2} \frac{1}{d(u)} & \text{if }v \in C_1 \cap C_2,\\
\frac{1}{d(v_0)} \sum_{u\in C_1} \frac{1}{d(u)} & \text{if }v \in C_1 \setminus C_2,\\
\frac{1}{d(v_0)} \sum_{u\in C_2} \frac{1}{d(u)} & \text{if }v \in C_2 \setminus C_1,\\
0 & \text{otherwise}.
\end{cases}
\end{split}
\]

Thus, if \(\alpha_2 = \frac{|C_2 \setminus C_1|}{|C_2|}\):
\[ \mathbf{p}(v) =
\begin{cases}
\frac{1}{|C_1\cup C_2|} (1, \alpha_1 + \alpha_2 + \beta) \overset{def} {=} \mathbf{p}_A & \text{if }v \in C_1\cap C_2,\\
\frac{1}{|C_1\cup C_2|} (1, \alpha_1 + \beta) \overset{def} {=} \mathbf{p}_B & \text{if }v \in C_1 \setminus C_2,\\
\frac{1}{|C_1\cup C_2|} (1, \alpha_2 + \beta) \overset{def} {=} \mathbf{p}_C & \text{if }v \in C_2 \setminus C_1,\\
0 & \text{otherwise}.
\end{cases}
\]

\subsection*{General case}\label{general-case}

In the general case where there may be more than one seed node,
we get:
 \[ \mathbf{p}(v) =
\begin{cases}
a \mathbf{p}_1 + b \mathbf{p}_A \overset{def} {=} \mathbf{p}_{1\cap 2} & \text{if }v \in C_1\cap C_2,\\
a \mathbf{p}_1 + b \mathbf{p}_B \overset{def} {=} \mathbf{p}_{1\setminus 2} & \text{if }v \in C_1 \setminus C_2,\\
a \mathbf{p}_2 + b \mathbf{p}_C \overset{def} {=} \mathbf{p}_{2\setminus 1} & \text{if }v \in C_2 \setminus C_1,\\
0 & \text{otherwise},
\end{cases}
\]
where  \(a=|S\cap (C_1 \setminus C_2)|\) and
\(b=|S\cap (C_1 \cap C_2)|\).
Observe that $\mathbf{p}_{1 \cap 2} \succ \mathbf{p}_{1\setminus 2} \succ \mathbf{p}_{2\setminus 1}$ for the lexicographical order.
In addition, the distances between  points in the embedding space are given by:
\[
\begin{split}
d_1 &= \lVert \mathbf{p}_{1\setminus 2} - \mathbf{p}_{1 \cap 2} \rVert = \frac{b}{|C_1\cup C_2|} \alpha_2,\\
d_2 &= \min(\lVert \mathbf{p}_{2\setminus 1} - \mathbf{p}_{1 \setminus 2} \rVert,
\lVert \mathbf{p}_{2 \setminus 1} - \mathbf{p}_{1 \cap 2} \rVert,\\
&= \frac{a}{|C_1|} \sqrt{1 + \alpha_1^2} + \frac{b}{|C_1\cup C_2|} |\alpha_1 - \alpha_2|.
\end{split}
\]
In particular, we have  \(d_1 < d_2\) if \(a/b\) is large enough.

\section{Noisy cliques}
\label{appendix-noisy-cliques}

Let \(S \subset C_1 \setminus C_2\) and  $\bar{\mathbf{p}}(v)=(\bar p_1(v),\bar p_2(v))$ be the embedding of the two-clique graph. We have:
\begin{align*}
p_1(v) &= \bar{p_1}(v) + \epsilon_1^A(v) - \epsilon_1^B(v),\\
p_2(v) &= \bar{p_2}(v) + \epsilon_2^A(v) - \epsilon_2^B(v),
\end{align*}
with: \[
\begin{split}
\epsilon_1^A(v) &=
\sum_{v_0 \in S: (v,v_0) \in E} \left(\frac{1}{d(v_0)} - \frac{1}{|C_1|}\right),\\
\epsilon_1^B(v) &=
\sum_{v_0 \in S: (v,v_0) \notin E} \frac{1}{|C_1|},\\
\epsilon_2^A(v) &=
\sum_{v_0 \in S} \sum_{\underset{(v,u)\in E, (u, v_0)\in E}{u\in C_1 \setminus C_2}} \left( \frac{1}{d(u)d(v_0)} - \frac{1}{|C_1|^2}\right)\\
&+ \sum_{v_0 \in S} \sum_{\underset{(v,u)\in E, (u, v_0)\in E}{u\in C_1 \cap C_2}} \left( \frac{1}{d(u)d(v_0)} - \frac{1}{|C_1||C_1 \cup C_2|}\right),\\
\epsilon_2^B(v) &=
\sum_{v_0 \in S} \sum_{\underset{(v,u)\notin E \text{ or } (u, v_0)\notin E}{u\in C_1 \setminus C_2}} \frac{1}{|C_1|^2} \\
&+ \sum_{v_0 \in S} \sum_{\underset{(v,u)\notin E \text{ or } (u, v_0)\notin E}{u\in C_1 \cap C_2}} \frac{1}{|C_1||C_1 \cup C_2|}.\\
\end{split}
\]

We have \(\epsilon_1^B(v) \neq 0\) for at most \(k\) nodes, and: \[
\begin{split}
\epsilon_1^A(v) & \leq
|S| \left( \frac{1}{|C_1| - k} - \frac{1}{|C_1|}\right) = \frac{k|S|}{|C_1|(|C_1| - k)}, \\
\epsilon_2^A(v) &=
|S| \Bigg[|C_1 \setminus C_2| \left( \frac{1}{(|C_1| - k)^2}  - \frac{1}{|C_1|^2} \right)\\
+
|C_1 \cap C_2| & \left( \frac{1}{(|C_1| - k)(|C_1 \cup C_2| - k)}  - \frac{1}{|C_1||C_1\cup C_2|)} \right) \Bigg],\\
\epsilon_2^B(v) & \leq
\frac{k}{|C_1|^2}.
\end{split}
\]

If \(\epsilon_1^B(v) = 0\), vectors
\(\epsilon(v) = (\epsilon_1^A(v) - \epsilon_1^B(v), \epsilon_2^A(v) - \epsilon_2^B(v))\)
satisfy
\(\lVert \epsilon(v) \rVert < \frac{|S|}{|C_1|} \sqrt{1 + \alpha_1^2} = d_2\)
provided  \(|C_1| / k\) is large enough.

\section{Adding external links}
\label{appendix-adding-external-links}

We have for all $v\in C_1 \cup C_2$:
\begin{align*}
p_1(v) &= \bar{p_1}(v) - \epsilon_1^{\rm in}(v),\\
p_2(v) &= \bar{p_2}(v) - \epsilon_2^{\rm in}(v),
\end{align*}
 and for all $v \notin C_1 \cup C_2$,
$p_1(v) = \epsilon_1^{\rm out}(v),
p_2(v) = \epsilon_2^{\rm out}(v),$
with
\[
\begin{split}
\epsilon_1^{\rm in}(v) & \leq
|S| \left( \frac{1}{|C_1|} - \frac{1}{|C_1| + l}\right) = \frac{l|S|}{|C_1|(|C_1| + l)}, \\
\epsilon_1^{\rm out}(v) & \leq
\frac{l}{|C_1|}, \\
\epsilon_2^{\rm in}(v) &\leq
|S| \Bigg[ |C_1 \setminus C_2| \left( \frac{1}{|C_1|^2} - \frac{1}{(|C_1| + l)^2} \right)\\
+
|C_1 \cap C_2| & \left( \frac{1}{|C_1||C_1\cup C_2|)} - \frac{1}{(|C_1| + l)(|C_1 \cup C_2| + l)}\right) \Bigg],\\
\epsilon_2^{\rm out}(v) & \leq
\frac{l}{|C_1|^2},
\end{split}
\]
We have  \(\lVert \epsilon^{\rm in}(v) \rVert < d_2\) and
\[\epsilon_2^{\rm out}(v) < \min((\mathbf{p}_{1\cap 2})_2,(\mathbf{p}_{1\setminus 2})_2,(\mathbf{p}_{2\setminus 1})_2) - d\]
provided \(|C_1|/l \) is large enough.

\section{References}

\hypertarget{refs}{}
\noindent\hypertarget{ref-andersen_communities_2006}{}
{[}1{]} Andersen, R. and Lang, K.J. 2006. Communities from seed sets.
\emph{Proceedings of the 15th international conference on World Wide
Web} (2006), 223--232.

\noindent\hypertarget{ref-andersen_local_2006}{}
{[}2{]} Andersen, R. et al. 2006. Local graph partitioning using
pagerank vectors. \emph{Foundations of Computer Science, 2006. FOCS'06.
47th Annual IEEE Symposium on} (2006), 475--486.

\noindent\hypertarget{ref-blondel_fast_2008}{}
{[}3{]} Blondel, V.D. et al. 2008. Fast unfolding of communities in
large networks. \emph{Journal of statistical mechanics: theory and
experiment}. 2008, 10 (2008), P10008.

\noindent\hypertarget{ref-chang_relative_2015}{}
{[}4{]} Chang, C.-S. et al. 2015. Relative centrality and local
community detection. \emph{Network Science}. (2015).

\noindent\hypertarget{ref-clauset_finding_2005}{}
{[}5{]} Clauset, A. 2005. Finding local community structure in networks.
\emph{Physical review E}. 72, 2 (2005).

\noindent\hypertarget{ref-ester_density-based_1996}{}
{[}6{]} Ester, M. et al. 1996. A density-based algorithm for discovering
clusters in large spatial databases with noise. \emph{Kdd} (1996).

\noindent\hypertarget{ref-jbpmm}{}
{[}7{]} Jeub, L.G. et al. 2015. Think locally, act locally: Detection of
small, medium-sized, and large communities in large networks.
\emph{Physical Review E}. 91, 1 (2015).

\noindent\hypertarget{ref-kloster_heat_2014}{}
{[}8{]} Kloster, K. and Gleich, D.F. 2014. Heat kernel based community
detection. \emph{Proceedings of the 20th ACM SIG- KDD international
conference on Knowledge discovery and data mining} (2014).

\noindent\hypertarget{ref-kloumann_community_2014}{}
{[}9{]} Kloumann, I.M. and Kleinberg, J.M. 2014. Community membership
identification from small seed sets. \emph{Proceedings of the 20th ACM
SIGKDD international conference on Knowledge discovery and data mining}
(2014).

\noindent\hypertarget{ref-lancichinetti_detecting_2009}{}
{[}10{]} Lancichinetti, A. et al. 2009. Detecting the overlapping and
hierarchical community structure in complex networks. \emph{New Journal
of Physics}. 11, 3 (2009).

\noindent\hypertarget{ref-lancichinetti_finding_2011}{}
{[}11{]} Lancichinetti, A. et al. 2011. Finding statistically
significant communities in networks. \emph{PloS one}. 6, 4 (2011).

\noindent\hypertarget{ref-li_uncovering_2015}{}
{[}12{]} Li, Y. et al. 2015. Uncovering the small community structure in
large networks: A local spectral approach. \emph{Proceedings of the 24th
international conference on world wide web} (2015).

\noindent\hypertarget{ref-mehler_expanding_2009}{}
{[}13{]} Mehler, A. and Skiena, S. 2009. Expanding network communities
from representative examples. \emph{ACM Transactions on Knowledge
Discovery from Data (TKDD)}. 3, 2 (2009).

\noindent\hypertarget{ref-newman_modularity_2006}{}
{[}14{]} Newman, M.E. 2006. Modularity and community structure in
networks. \emph{Proceedings of the national academy of sciences}. 103,
23 (2006).

\noindent\hypertarget{ref-pons_computing_2005}{}
{[}15{]} Pons, P. and Latapy, M. 2005. Computing communities in large
networks using random walks. \emph{International Symposium on Computer
and Information Sciences} (2005).

\noindent\hypertarget{ref-reid_partitioning_2013}{}
{[}16{]} Reid, F. et al. Partitioning breaks communities.
\emph{Mining Social Networks and Security Informatics}. Springer.
(2013).

\noindent\hypertarget{ref-rosvall_maps_2008}{}
{[}17{]} Rosvall, M. and Bergstrom, C.T. 2008. Maps of random walks on
complex networks reveal community structure. \emph{Proceedings of the
National Academy of Sciences}. 105, 4 (2008).

\noindent\hypertarget{ref-spielman_local_2013}{}
{[}18{]} Spielman, D.A. and Teng, S.-H. 2013. A local clustering
algorithm for massive graphs and its application to nearly linear time
graph partitioning. \emph{SIAM Journal on Computing}. 42, 1 (2013).

\noindent\noindent\hypertarget{ref-yang_defining_2015}{}
{[}19{]} Yang, J. and Leskovec, J. 2015. Defining and evaluating network
communities based on ground-truth. \emph{Knowledge and Information
Systems}. 42, 1 (2015).

\end{document}